\documentclass[cameraready]{Interspeech}

\title{ARTT: Augmented Reverberant-Target Training for Unsupervised Monaural Speech Dereverberation}

\author[affiliation={}]{Siqi}{Song}
\author[affiliation={}]{Fulin}{Wu}
\author[affiliation={}]{Zhong-Qiu}{Wang}

\address{
    Department of Computer Science and Engineering,\\Southern University of Science and Technology, Shenzhen, China
}

\email{siqi.song1@outlook.com, wang.zhongqiu41@gmail.com}


\usepackage{comment}
\usepackage{multirow}
\usepackage{microtype}
\usepackage{pifont}
\newcommand{\cmark}{\ding{51}} 
\newcommand{\xmark}{\ding{55}} 
\usepackage{hyperref}

\usepackage{cite}
\usepackage{flushend}
\usepackage{caption}
\begin{document}

\maketitle

\begin{abstract}
Due to the absence of clean reference signals and spatial cues, monaural unsupervised speech dereverberation is a challenging ill-posed inverse problem. 
To realize it, we propose augmented reverberant-target training (ARTT), which consists of two stages.
In the first stage, reverberant-target training (RTT) is proposed to first further reverberate the observed reverberant mixture signal, and then train a deep neural network (DNN) to recover the observed reverberant mixture via discriminative training.
Although the target signal to fit is reverberant, we find that the resulting DNN can effectively reduce reverberation.
In the second stage, an online self-distillation mechanism based on the mean-teacher algorithm is proposed to further improve dereverberation.
Evaluation results demonstrate that ARTT achieves strong unsupervised dereverberation performance, significantly outperforming previous baselines.
\end{abstract}

\begin{figure*}[t]
  \centering
  \makebox[\textwidth][c]{
    \hspace{0.5cm} 
    \includegraphics[width=1.15\textwidth]{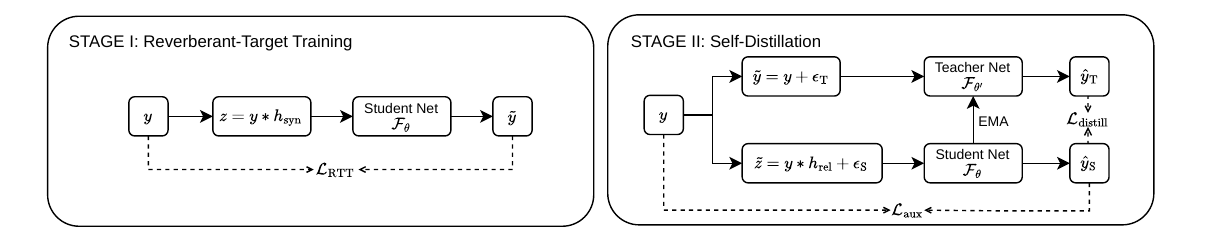}
  }
  \vspace{-0.6cm}
  \caption{ARTT overview.}
  \label{fig:ARTT}
\end{figure*}

\section{Introduction}

Room reverberation is a ubiquitous phenomenon in real-world acoustic environments \cite{naylor2010speech,kuttruff2016room,A.P.Habets2018}. While natural to human ears, it severely degrades speech intelligibility and impairs downstream applications such as automatic speech recognition and speaker recognition \cite{yoshioka2012making, kinoshita2016summary}.
Although multi-microphone systems can exploit spatial diversity to mitigate these effects, monaural blind dereverberation remains a fundamentally ill-posed inverse problem \cite{wang2024usdnet}. It requires decoupling the anechoic speech and room impulse response (RIR) based on only a single-channel noisy-reverberant mixture signal, without access to reference signals for model training or spatial cues for suppressing reverberation.

Supervised deep learning effectively trains deep neural networks (DNNs) to map reverberant inputs to anechoic targets \cite{WDLreview, hu2020dccrn, wang2023tfgrid, kimura2024diffusion}. However, a fundamental constraint is the requirement of pairs of reverberant mixture signals and their corresponding anechoic speech for supervised training.
Since acquiring the clean anechoic signals of real-recorded reverberant mixture signals is often difficult, models are typically trained on synthetic datasets created by using a room simulator \cite{allen1979image, scheibler2018pyroomacoustics}.
This, however, often leads to a domain mismatch problem between model training and testing, causing performance degradation when the trained models are applied to unseen acoustic environments \cite{kinoshita2016summary}.

To overcome this, early unsupervised approaches, such as weighted prediction error (WPE) \cite{nakatani2010wpe, jukic2015wpe, yoshioka2012wpe}, employ delayed linear prediction to estimate late reverberation, which is then subtracted from the mixture for dereverberation. While WPE is computationally efficient and effective for suppressing late reverberation, it does not fully exploit the rich speech priors and room acoustics that data-driven approaches can leverage.

Neural methods like USDnet \cite{wang2024usdnet} incorporate physical consistency by optimizing mixture-constraint losses. This mechanism trains DNNs solely on a training set of reverberant mixtures by constraining that the network output (i.e., the estimated clean source), if linearly filtered in a proper way, should reconstruct the observed reverberant mixture. To obtain the filters, forward convolutive prediction \cite{wang2021fcp} is used to estimate the linear filters that best align the network output with the mixture. However, while USDnet enforces physical constraints, it typically requires multi-channel mixture-constraint losses to ensure sufficiently robust constraints and accurate dereverberation.
Recent generative methods such as BUDDy \cite{lemercier2025buddy} and USD-DPS \cite{wu2025unsupervised} extend the diffusion framework to handle unknown RIRs. By including a parametric RIR model, they perform joint room acoustics estimation and dereverberation along reverse diffusion trajectory. Despite their generalization ability, iterative joint estimation incurs a heavy computational burden.

There are alternative unsupervised methods bypassing filter computation entirely \cite{fu2022metricgan, tzinis2022remixit}. The noisy-target training (NyTT) framework \cite{fujimura2021nytt}, based on the Noise2Noise principle \cite{lehtinen2018noise2noise}, first adds noises to each observed noisy mixture and then trains a DNN, in a discriminative fashion, to reconstruct the original noisy mixture based on the newly-created noisy mixture.
Since the DNN cannot predict random independent noise, optimizing the reconstruction loss encourages the model to learn the shared underlying speech patterns and realize denoising.
However, standard NyTT fundamentally struggles with speech dereverberation, as room reverberation is a highly-correlated convolutional process, which violates the independent noise assumption.

In this context, we propose augmented reverberant-target training (ARTT), an unsupervised approach for monaural unsupervised speech dereverberation. Inspired by NyTT \cite{fujimura2025nytt}, we propose reverberant-target training (RTT), where each observed reverberant mixture signal is first convolved with a stochastic synthetic impulse response to add more reverberation, and then a DNN is trained in a discriminative way to recover the observed reverberant signal based on the newly-created mixture. Although effective, RTT is found not stable and has limited performance.
To address these issues, inspired by a teacher-student learning algorithm named BYOL~\cite{grill2020bootstrap} we propose to augment RTT with a continuous online self-distillation mechanism, which utilizes an exponential moving average (EMA) teacher to dynamically generate consistent targets for teacher-student learning.
In addition, by corrupting the student's input with both synthetic reverberation and noise, we achieve dereverberation and denoising within a unified, stable loop.
Evaluation results demonstrate that ARTT outperforms existing monaural unsupervised methods.
A sound demo is provided in the link below\footnote{See \url{https://arttdemo.github.io/artt_demo/}.}.

\section{ARTT}

ARTT deals with monaural speech dereverberation.
Let $x$ denote direct-path signal, $h$ the corresponding, unknown relative transfer function (RTF) relating the direct-path signal to reverberant speech, and $v$ the additive background noise, the observed mixture signal $y$ can be formulated, in the time domain, as follows:
\begin{align}
y = x \ast h + v,
\end{align}
where $\ast$ denotes linear convolution.
In unsupervised speech dereverberation, we assume that we only have access to a dataset $\mathcal{D} = \{y_i\}_{i=1}^N$ consisting of $N$ observed reverberant mixture signals, and the clean source $x$ and RTF $h$ are both inaccessible.
Our goal is to train an unsupervised DNN model $\mathcal{F}_\theta$ that can estimate the direct-path signal, $\hat{x} = \mathcal{F}_\theta(y)$, solely relying on $\mathcal{D}$.
The proposed ARTT approach consists of two stages, reverberant-target training and self-distillation, both of which are described next.
See Fig. \ref{fig:ARTT} for an overview.

\subsection{Stage I: Reverberant-Target Training}\label{ssec:StageI}

To deal with the fact that clean target speech is not available for model training in an unsupervised setup, we propose reverberant-target training (RTT) to enable discriminative training and realize initial unsupervised dereverberation.
The idea is to train a supervised, discriminative DNN model to reconstruct the observed reverberant mixture signal $y$ from a further reverberated mixture $z = y \ast h_\text{syn}$, where $h_\text{syn}$ is a synthetic statistical RTF to be described later in Eq. (\ref{statistical_RIR_eq}).
Since the input $z$ is $x \ast h \ast h_\text{syn}$ and due to the commutative property of linear convolution, the model would not be able to learn to surgically remove $h_\text{syn}$ while maintaining $h$.
Intuitively, it would tend to learn to reduce both $h_\text{syn}$ and $h$, thereby to some extent realizing dereverberation.

\begin{figure}[h]
    \centering
    \includegraphics[width=0.9\linewidth]{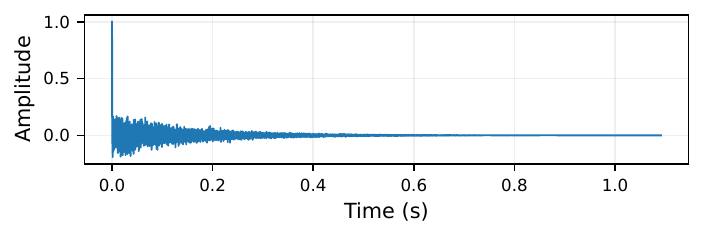}
    \vspace{-0.4cm}
    \caption{Example synthetic statistical RTF $h_\text{syn}$ for RTT.}\label{fig:rir}
    \vspace{-0.3cm}
\end{figure}

Following a stochastic reverberation model \cite{habets2010speech}, the synthetic statistical RTF $h_\text{syn}$ is designed to have a unit impulse
at the first sample followed by an exponentially decaying diffuse tail, as shown in Fig. \ref{fig:rir}.
In detail, it is modeled as:
\begin{equation}
h_\text{syn}[n] = \left\{
\begin{aligned}
&1,\,\,\text{for}\,\,n = 0; \\
&\gamma \cdot \eta[n] \cdot e^{-\lambda\cdot n},\,\,\text{for}\,\,n > 0; \\
& 0,\,\, \text{otherwise},
\end{aligned}
\right.
\label{statistical_RIR_eq}
\end{equation}
where
$\eta[n] \sim \mathcal{N}(0, 1)$ is sampled from Gaussian white noise, $\lambda$ determines the decay rate corresponding to a randomly sampled reverberation time ($T_{60}$), and $\gamma$ is a scaling factor controlling the energy of late reverberation relative to that of the first sample.

The DNN model takes as input the reverberated mixture $z = y \ast h_\text{syn}$, and is trained to reconstruct the original observed mixture signal $y$ via supervised training using the following loss:
\begin{equation}
    \mathcal{L}_\text{RTT} = \mathcal{L}_\text{rec}\big(\mathcal{F}_\theta(y \ast h_\text{syn}), y\big),
\end{equation}
where $\mathcal{L}_\text{rec}(\cdot,\cdot)$ is a distance function to be described later in Eq. (\ref{eq:loss_rec}).
In this stage, RTT enables the model to learn how to suppress the synthetic late reverberation, providing an initial
dereverberation result for subsequent self-distillation.

\subsection{Stage II: Self-Distillation}\label{ssec:StageII}

Although in our experiments RTT is effective at dereverberation, we also observe that it has limited performance at isolating direct-path signal, and in addition the network risks overfitting to the observed mixture $y$.
To improve the performance, we adapt the mean-teacher framework \cite{tarvainen2017mean} to the task of speech dereverberation, where the teacher network's parameters $\theta'$ are updated via an EMA of the student network's parameters.
In detail, the teacher parameters $\theta'$ are updated iteratively from the student parameters $\theta$ at training step $k$ using the EMA strategy:
\begin{equation}
    \theta'_k = \alpha \cdot \theta'_{k-1} + (1 - \alpha) \cdot \theta_k,
\end{equation}
where $\alpha \in [0, 1)$ is a momentum factor controlling the update smoothness. 
In essence, the teacher network serves as a temporal ensemble of the student, and could provide coherent and stable pseudo-labels for improving the student.
Specifically, we design an asymmetric training strategy where the student and teacher process differently-corrupted signals that naturally drives the model to suppress reverberation.

\textbf{\textit{1) Augmented Input Construction}}:
First, a physically-simulated %
RTF $h_\text{rel}$, which is acoustically more realistic than $h_\text{syn}$ in Eq. (\ref{statistical_RIR_eq}), is created to refine the initial mapping learned in Stage I.
It is computed by first using a room simulator to generate a full reverberant RIR $h_\text{sim}$ and the corresponding direct-path component $h_\text{dir}$, and then using the following method:
\begin{equation}
    h_\text{rel} = \texttt{iFFT} \left( \frac{\texttt{FFT}(h_\text{sim})}{\texttt{FFT}(h_\text{dir})} \right),
    \label{eq:deconv}
\end{equation}
where $\texttt{FFT}$ and $\texttt{iFFT}$ respectively denote the fast Fourier transform and its inverse. Note that $h_\text{rel}$ serves as a correction filter that maps the direct sound to the reverberant observation in the simulator's acoustic space.
To force the network to learn a noise-invariant representation and implicitly denoise the observation $y$, we further inject independent Gaussian noises $\epsilon_\text{T} \sim \mathcal{N}(0, \sigma_\text{T}^2)$ and $\epsilon_\text{S} \sim \mathcal{N}(0, \sigma_\text{S}^2)$, where the subscripts ``T'' and ``S'' respectively mean teacher and student. This results in two distinct inputs:
\begin{itemize}
    \item \textit{Teacher Input (Stable):} $\tilde{y} = y + \epsilon_\text{T}$, which serves as the pseudo-target generator.
    \item \textit{Student Input (Hard):} $\tilde{z} = y \ast h_\text{rel} + \epsilon_\text{S}$, which contains both synthetic reverberation and noise perturbation.
\end{itemize}
This construction creates an \textit{asymmetric} learning task. The teacher network receives a cleaner input (i.e., $\tilde{y}$) to maintain reliable guidance, while the student network is exposed to a heavily corrupted input (i.e., $\tilde{z}$) augmented with both the relative reverberation and additive noise.
This forces the student network to implicitly invert the RTF $h_\text{rel}$ and denoise the signal to match the teacher network's canonical representation.

\textbf{\textit{2) Distillation Objectives}}:
The goal is to align the student's prediction $\mathcal{F}_\theta(\tilde{z})$ with the teacher's stable target $\hat{y}_\text{T} = \mathcal{F}_{\theta'}(\tilde{y})$.
The primary self-distillation loss enforces invariance against the synthetic reverberation and noise:
\begin{equation}
    \mathcal{L}_\text{distill} = \mathcal{L}_\text{rec}\big(\mathcal{F}_\theta(\tilde{z}), \texttt{SG}(\hat{y}_\text{T})\big),
\end{equation}
where $\texttt{SG}(\cdot)$ stands for the stop-gradient operator.
To avoid convergence to trivial solutions, we introduce an auxiliary regularization term that utilizes the original observation $y$ as a reference constraint:
\begin{equation}
    \mathcal{L}_\text{aux} = \mathcal{L}_\text{rec}\big(\mathcal{F}_{\theta}(\tilde{z}), y\big).
\end{equation}
The overall objective combines the above two loss functions using a weighting term $\omega$ ($>0$):
\begin{equation}\label{loss_stage_II}
    \mathcal{L}_\text{StageII} = \mathcal{L}_\text{distill} + \omega \cdot \mathcal{L}_\text{aux}.
\end{equation}

\subsection{Loss Function}
We employ a unified reconstruction loss function $\mathcal{L}_\text{rec}(\hat{u}, u)$ across both training stages, with $u$ denoting the reference target signal (specifically, $y$ in~\hyperref[ssec:StageI]{Stage I} and $\hat{y}_\text{T}$ in~\hyperref[ssec:StageII]{Stage II}) and $\hat{u}$ the estimated signal. It combines time- and time-frequency-domain losses:
\begin{equation}
    \mathcal{L}_\text{rec}(\hat{u}, u) = \mathcal{L}_{\mathrm{SI-SDR-SE}}(\hat{u}, u) + \mathcal{L}_{\mathrm{Mag}}(\hat{u}, u),
    \label{eq:loss_rec}
\end{equation}
which are respectively defined as follows:
\begin{align}
    \mathcal{L}_{\mathrm{SI\text{-}SDR\text{-}SE}}
        &= -10 \cdot \log_{10} \frac{\lVert u \rVert_2^2}{\lVert \hat{\beta} \cdot \hat{u} - u \rVert_2^2},\,\text{with}\, \hat\beta
        = \frac{\hat{u}^\mathsf{T} u}{\hat{u}^\mathsf{T} \hat{u}}, \\
    \mathcal{L}_{\mathrm{Mag}} &= \frac{1}{T \times F} \left\lVert \big| \mathrm{STFT}(\hat{u}) \big| - \big| \mathrm{STFT}(u) \big| \right\rVert_1.
\end{align}

\section{Experimental Setup}

\subsection{Dataset and Evaluation Metrics}

We validate the proposed algorithms based on the WSJ0CAM-DEREVERB dataset, a popular benchmark widely used in dereverberation studies \cite{wang2021fcp, wang2024usdnet, wu2025unsupervised}.
The dry source signals are from the WSJ0CAM corpus, which has $7,861$, $742$ and $1,088$ utterances in its training, validation and test sets, respectively.
Based on them, $39,293$ ($\sim$$77.7$ h), $2,968$ ($\sim$$5.6$ h) and $3,262$ ($\sim$$6.4$ h) noisy-reverberant mixtures are respectively simulated as the training, validation and test sets.
An 8-channel circular microphone array (with a diameter of $20$~cm) records speech at distances between $0.75$ and $2.5$~m, with reverberation times ($T_{60}$) between $0.2$ and $1.3$~s. Diffuse air-conditioning noise from the REVERB dataset~\cite{kinoshita2016summary} is added at randomly-chosen SNRs from $5$ to $25$~dB.
The sampling rate is $16$ kHz. We only use the first channel of the dataset for training and testing.

We use the direct-path signal at the first channel as the reference signal for metric computation. We report perceptual evaluation of speech quality (PESQ)~\cite{rix2001perceptual} scores to measure perceptual quality, extended short-time objective intelligibility (eSTOI)~\cite{jensen2016algorithm} to evaluate speech intelligibility, and SI-SDR~\cite{le2019sdr} to evaluate the accuracy of dereverberated signals at the sample level.
We use the \texttt{python-pesq} toolkit to compute narrow-band PESQ, and the \texttt{pystoi} toolkit to compute eSTOI.

\subsection{Method Configuration}

For RTT, to prevent overfitting and ensure robustness, we implement it by generating synthetic RIRs $h_\text{syn}$ online during training. The reverberation time ($T_{60}$) is sampled from a uniform distribution $\mathcal{U}(0.5, 1.2)$ s, while the direct-to-reverberant energy ratio (DRR) is drawn from $\mathcal{U}(-16, -6)$ dB. The stochastic diffuse tail is generated via Gaussian white noise modulated by an exponentially-decaying envelope, with the scaling factor $\gamma$ dynamically computed to ensure its energy relative to the unit direct-path matches the sampled DRR.

For self-distillation, the \texttt{pyroomacoustics} toolkit~\cite{scheibler2018pyroomacoustics} is employed to simulate physically-consistent environments using the image method. Room dimensions are uniformly sampled with length and width from the range $[5.0, 10.0]$ m, and height from $[3.0, 4.0]$ m. The $T_{60}$ ranges from $0.2$ to $1.3$ s. The relative RIR $h_\text{rel}$ is derived via spectral deconvolution using Eq.~\eqref{eq:deconv}.
The weighting term in Eq. (\ref{loss_stage_II}) for the auxiliary loss is set to $\omega=1.2$. The noise standard deviations $\sigma_\text{T}$ and $\sigma_\text{S}$ are set to $0.02$ times the standard deviation of the observation $y$. The momentum factor $\alpha$ is set to $0.999$.

We use TF-GridNet \cite{wang2023tfgrid} as our DNN architecture. We set its hyper-parameters to $D=128, B=4, I=1, J=1, H=200, L=3$ and $E=4$. It is trained via complex spectral mapping \cite{Tan2020CSM,wang2020multi,Wang2020CSM,Wang2021CSMSS} to predict the real and imaginary (RI) components of the target signals from the RI components of the input mixtures.
For STFT, the window size is 32 ms, hop size 8 ms, and the square root of the Hann window is used as the analysis window.

\subsection{Baselines}

We compare our method with unsupervised baselines including WPE~\cite{nakatani2010wpe}, USDnet~\cite{wang2024usdnet} and BUDDy~\cite{lemercier2025buddy}, as well as a supervised baseline, DNN-WPE~\cite{kinoshita2017neural}. 
For WPE, we use the implementation in the \texttt{nara-wpe} toolkit~\cite{drude2018nara}, and the STFT configuration is consistent with our experiment. The filter tap length is set to $37$,
the prediction delay to $3$, and the algorithm is run for $3$ iterations. 
For USDnet, we report their best results in row 2a in Table II and 3a in Table VIII of the original paper~\cite{wang2024usdnet}. 
For BUDDy, we report the results of their best setup.
It is based on the 1D U-Net variant, which outperforms the original NCSN++ version in Table 1 of~\cite{wu2025unsupervised}. For DNN-WPE~\cite{kinoshita2017neural}, we use the same DNN in our framework, and the STFT configuration follows the WPE baseline.

\begin{table*}[h]
\footnotesize
    \centering
    \caption{Results on test set of WSJ0CAM-DEREVERB.}
    \label{tab:comparison_results}
    
    \setlength{\tabcolsep}{9pt} 
    
    \begin{tabular}{cl c cc ccc}
        \toprule
        \raisebox{-1.5ex}[0pt][0pt]{Row} & 
        \raisebox{-1.5ex}[0pt][0pt]{System} & 
        \raisebox{-1.5ex}[0pt][0pt]{Unsupervised} &
        \multicolumn{2}{c}{Number of Channels in} & 
        \raisebox{-1.5ex}[0pt][0pt]{SI-SDR (dB)$\uparrow$} & 
        \raisebox{-1.5ex}[0pt][0pt]{PESQ$\uparrow$} & 
        \raisebox{-1.5ex}[0pt][0pt]{eSTOI$\uparrow$} \\
        
        \cmidrule(lr){4-5} 
        
        & & & Input & Loss & & & \\
         
        \midrule
        0a & Mixture & - & $1$ & - & $-3.6$ & $1.64$ & $0.494$ \\
        \midrule
        1a & WPE \cite{nakatani2010wpe} & \cmark & $1$ & - & $-1.7$ & $1.78$ & $0.529$ \\
        1b & WPE \cite{nakatani2010wpe} & \cmark & $8$ & - & \phantom{$-$}$1.3$ & $2.02$ & $0.690$ \\
        \midrule
        2a & USDnet \cite{wang2024usdnet} & \cmark & $1$ & $1$ & $-2.1$ & $1.76$ & $0.561$ \\
        2b & USDnet \cite{wang2024usdnet} & \cmark & $1$ & $8$ & \phantom{$-$}$2.9$  & $2.53$ & $0.772$ \\
        2c & BUDDy \cite{lemercier2025buddy} & \cmark & $1$ & - & \phantom{$-$}$2.1$  & $2.49$ & $0.802$ \\
        \midrule
        3a & DNN-WPE \cite{kinoshita2017neural} & \xmark & $1$ & - & \phantom{$-$}$2.8$ & $2.16$ & $0.744$ \\
        \midrule
        4a & {ARTT\textsubscript{StageI} (RTT)} & \cmark & $1$ & - & \phantom{$-$}$3.3$  & $2.18$ & $0.740$ \\
        4b & {ARTT\textsubscript{StageII} (RTT+Self-Distillation)} & \cmark & $1$ & - & \phantom{$-$}$\mathbf{7.3}$  & $\mathbf{2.61}$ & $\mathbf{0.832}$ \\
        \bottomrule
    \end{tabular}
\end{table*}

\section{Evaluation Results}

Table~\ref{tab:comparison_results} reports the results on the test set of WSJ0CAM-DEREVERB. The unprocessed mixture suffers from severe reverberation, resulting in a low SI-SDR of $-3.6$ dB.
While traditional WPE and single-channel USDnet show limited improvements, the recent generative method BUDDy achieves a competitive SI-SDR of $2.1$ dB. In contrast, our proposed ARTT\textsubscript{StageI} model achieves an SI-SDR of $3.3$ dB, surpassing all unsupervised baselines and demonstrating the effectiveness of RTT.
Introducing Stage II provides a significant performance boost, further elevating the performance to state-of-the-art levels, reaching $7.3$ dB in SI-SDR and $2.61$ in PESQ.
We then compare our framework with the supervised baseline, DNN-WPE in row $3$a. Our unsupervised framework significantly outperforms it in all metrics.

\begin{table}
    \footnotesize
    \centering
    \caption{Ablation results of ARTT\textsubscript{StageII} on WSJ0CAM-DEREVERB.}
          \vspace{-0.2cm}
    \label{tab:ablation_stage2}
    \setlength{\tabcolsep}{6pt}
    \begin{tabular}{cc ccc}
        \toprule
        
        Inject Noise ($\epsilon$) & $\mathcal{L}_\text{aux}$ & SI-SDR (dB)$\uparrow$ & PESQ$\uparrow$ & eSTOI$\uparrow$ \\
        \midrule
        \cmark & \xmark & \phantom{$-$}$3.1$ & $1.97$ & $0.727$ \\
        \xmark & \cmark & \phantom{$-$}$6.0$ & $2.52$ & $0.820$ \\
        \cmark & \cmark & \phantom{$-$}$\mathbf{7.3}$ & $\mathbf{2.61}$ & $\mathbf{0.832}$ \\
        \bottomrule
    \end{tabular}
\end{table}

\begin{figure}[h]
    \centering
    \includegraphics[width=1.0\linewidth]{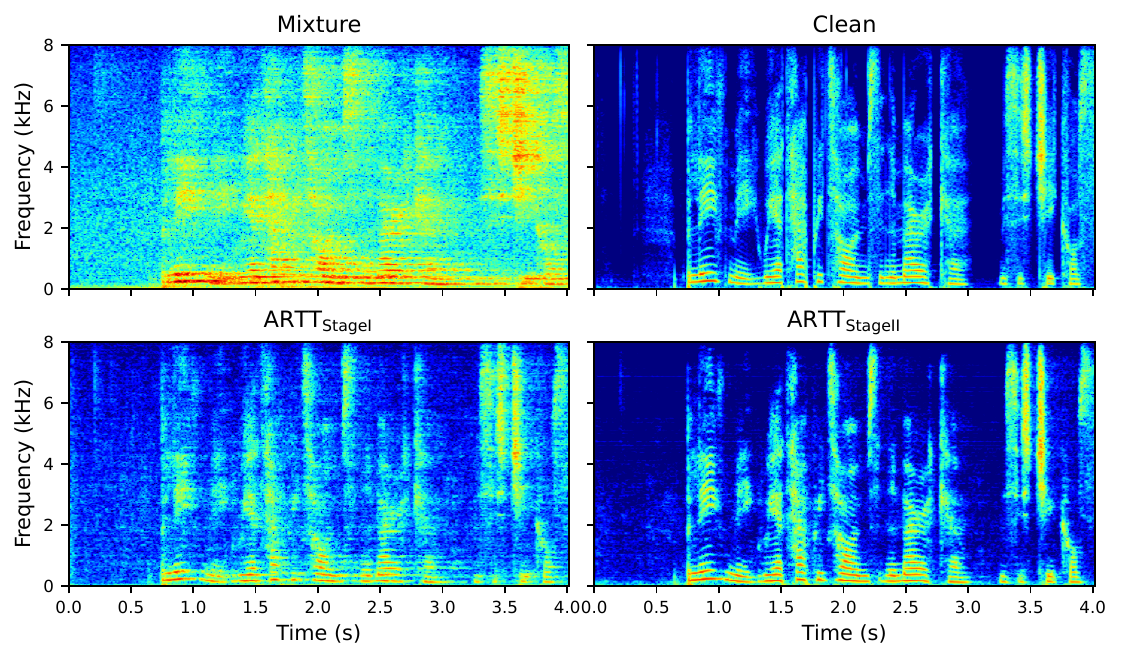}
      \vspace{-0.5cm}
    \caption{Spectrogram visualization of reverberant mixture, clean reference signals (direct-path), and outputs of ARTT\textsubscript{StageI} and ARTT\textsubscript{StageII}.}
    \label{fig:spectrogram}
\end{figure}

Table~\ref{tab:ablation_stage2} validates the contribution of each component in Stage II of ARTT.
Removing the auxiliary loss leads to a sharp degradation across all metrics. This confirms that, without the regularization from the reference signal, the self-distillation process becomes unstable and prone to model collapse.
Excluding the noise injection $\epsilon$ also results in a noticeable decline, particularly in SI-SDR. This indicates that the independent noise perturbation effectively guides the model to suppress the background interference present in the mixtures.

\begin{figure}
    \centering
    \includegraphics[width=1.0\linewidth]{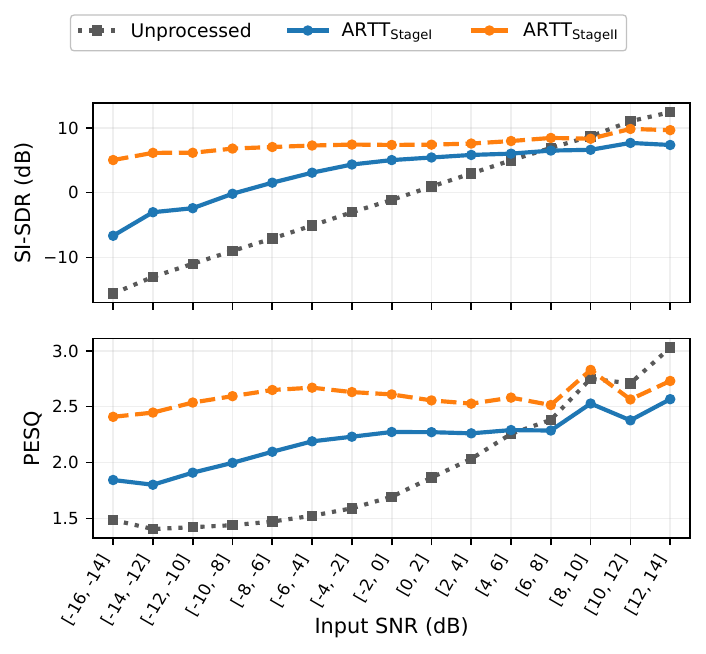}
          \vspace{-0.5cm}
    \caption{SI-SDR (dB) and PESQ comparison between ARTT\textsubscript{StageI} and ARTT\textsubscript{StageII} at different ranges of input SNR levels. Best viewed in color.}
    \label{fig:sisdr}
\end{figure}

Figure~\ref{fig:sisdr} evaluates the robustness of ARTT by comparing Stage I and II at different input SNR levels on the test set of WSJ0CAM-DEREVERB. 
With self-distillation, ARTT\textsubscript{StageII} demonstrates significantly higher stability, maintaining relatively stable performance at all the input SNR levels, while ARTT\textsubscript{StageI} degrades rapidly in low input SNR ranges. This confirms that the self-distillation process effectively rectifies the instability of the initial dereverberation by RTT in noisy environments.

In Figure~\ref{fig:spectrogram}, we show example outputs from the two stages of ARTT on a test mixture sampled from WSJ0CAM-DEREVERB.
We observe that ARTT\textsubscript{StageI} effectively suppresses late reverberation, significantly reducing the smearing effects observed in the mixture. However, the output appears slightly over-smoothed, with a loss of fine spectral textures.
In comparison, ARTT\textsubscript{StageII} better preserves the fine spectral structures and better reduces the background noise, exhibiting sharper harmonic structures and clearer formants that closely match the clean reference signal.

\section{Conclusion}

We have proposed ARTT, an unsupervised training framework for monaural blind dereverberation. RTT effectively addresses the challenge of missing clean references by constructing supervision signals directly from reverberant observations. In addition, the introduction of an online self-distillation stage exploiting asymmetric acoustic inputs significantly enhances the model's robustness against noise and artifacts.
Evaluation results on the WSJ0CAM-DEREVERB dataset demonstrate that ARTT outperforms existing unsupervised baselines.

\section{Generative AI Use Disclosure}
Generative AI was used only for language polishing. All scientific content and results were produced by the authors, who take full responsibility for the paper.

\bibliographystyle{IEEEtran}
\bibliography{mybib}

\end{document}